\documentclass[pre,twocolumn,amsmath,amssymb,nofootinbib]{revtex4}
\usepackage{graphicx,color,dcolumn,subfigure,bm}

\newlength\fwidth
\newlength\fheight

\begin{document}


\title{The optimal swimming sheet}
\author{Thomas D. Montenegro-Johnson}
\author{Eric Lauga}
\email{e.lauga@damtp.cam.ac.uk}
\affiliation{Department of Applied Mathematics and Theoretical Physics, Centre
for Mathematical Sciences, University of Cambridge, Wilberforce Road, Cambridge
CB3 0WA, UK}
\date{\today}
\begin{abstract}
Propulsion at microscopic scales is often achieved through propagating traveling
waves along hair-like organelles called flagella. Taylor's two-dimensional
swimming sheet model is frequently used to provide insight into problems of
flagellar propulsion. We derive numerically the large-amplitude waveform of the
two-dimensional swimming sheet that yields optimum hydrodynamic efficiency; the
ratio of the squared swimming speed to the rate-of-working of the sheet against
the fluid. Using the boundary element method, we show the optimal waveform is a
front-back symmetric regularized cusp that is $25\%$ more efficient than the
optimal sine-wave. This optimal two-dimensional shape is smooth, qualitatively
different from the kinked form of Lighthill's optimal three-dimensional
flagellum, not predicted by small-amplitude theory, and different from the
smooth circular-arc-like shape of active elastic filaments. 
\end{abstract} 

\keywords{Swimming sheet, Stokes flow, micro-scale propulsion, optimal waveform} 
\maketitle


\section{Introduction} 
\label{sec:introduction}

The microscopic world is teeming with organisms and cells that must self-propel
through their fluid environment in order to survive or carry out their functions
\cite{braybook}.  At these very small scales, viscous forces dominate inertia in
fluid flows, and a common method of overcoming the challenges of viscous
propulsion through the fluid environment is by propagating waves along slender,
hair-like organelles called  flagella or cilia
\cite{lighthill1975mathematical,childress1981mechanics}. To examine the fluid
mechanical basis for microscopic propulsion, \citet{taylor1951analysis}
considered a simplified flagellum model comprising a two-dimensional sheet
exhibiting small amplitude traveling waves. This seminal work subsequently
sparked the development of other techniques for examining Newtonian viscous
flows such as slender-body theory \cite{hancock1953self,keller76-jfm,johnson80}
and resistive-force theory \cite{gray1955propulsion,lighthill1975mathematical},
as well as other models for non-Newtonian swimming based on distribution of
force singularities  \cite{montenegro2012modelling,chrispell2013actuated} .

Due to its analytical tractability and agreement with more involved approaches
in the small-amplitude limit, Taylor's swimming sheet has been used to give
insight into many fundamental problems in microscale propulsion, such as
hydrodynamic synchronization between waving flagella
\cite{taylor1951analysis,elfring2009hydrodynamic,elfring2010two,elfring2011synchronization},
swimming in non-Newtonian fluids \cite{lauga2007propulsion,
teran2010viscoelastic, montenegro2013physics} and swimming past deformable
membranes \cite{chrispell2013actuated}. These approaches are typically
characterized by asymptotic expansion of the flagellar waveform under the
condition that the amplitude of the waves is small when compared to the
wavelength. Recently, Taylor's small-amplitude expansion was formally extended
to arbitrarily high order for a pure sine-wave, a method able to produce results
comparable to full numerical simulations of large amplitude sine-waves with the
boundary element method \cite{sauzade2011taylor}.

Motivated by the role of evolutionary pressures on the shape and kinematics of
swimming microorganisms, it is relevant to investigate which flagellar waveform
is the most energetically efficient for the cell. For an infinite flagellum,
\citet{lighthill1975mathematical} showed that in the local drag approximation of
resistive-force theory, the hydrodynamically-optimal flagellar waveform has a
constant tangent angle to the swimming direction. This leads to the shape of a
smooth helix in three dimensions, and a singular triangle wave in two
dimensions. Whilst the helical waveform is commonly observed in bacterial
flagella \cite{brennen77,Spagnolie2011}, unsurprisingly the kinked planar
waveform is not. \citet{spagnolie2010optimal} showed that this shape singularity
in Lighthill's flagellum can be regularized by penalizing the swimming
efficiency by the elastic energy required to bend a flagellum, which might
provide one explanation for its absence in nature. This model was then improved
upon by \citet{lauga2013shape} by proposing an energetic measure based on the
internal molecular cost necessary to deform the active flagellum. For
finite-length flagella, \citet{pironneau1974optimal} showed that traveling waves
are fundamental to optimal propulsion. Using resistive force theory, they then
analyzed optimal patterns for model spermatozoon exhibiting small amplitude planar sinusoidal waves and finite amplitude triangle waves. {The optimal stroke pattern of Purcell's finite three-link swimmer was found by
\citet{tam2007optimal}, who then went on to consider optimal gaits for the green
alga {\it Chlamydomonas} \cite{tam2011optimal}.}

While all past optimization work has focused on three-dimensional slender
filaments, the optimal waveform of Taylor's two-dimensional swimming sheet has
yet to be considered. Although the two-dimensionality of the problem makes it
less realistic as a  model for swimming cells, the fluid dynamics around a
sheet can be computed very accurately, allowing us to bypass various
hydrodynamic modeling approximations employed in three dimensions. In the
present study, we use the boundary element method to examine a sheet
propagating large amplitude waves of arbitrary shape. We derive computationally
the waveform leading to swimming with maximum hydrodynamic efficiency. We show
that the optimal waveform for the swimming sheet is a regularized cusp wave not
predicted by small-amplitude analysis. The optimal is qualitatively different
from three-dimensional swimmers, both the kinked triangle of Lighthill's
hydrodynamically-optimal flagellum \cite{lighthill1975mathematical} and the
circular arcs of internally-optimal active filaments \cite{lauga2013shape} and
indicates a qualitative difference between two- and three-dimensional swimming at
large amplitude.


\section{Formulation of the problem} 
\label{sec:modeling}

Newtonian fluid mechanics at microscopic scales is governed by the incompressible Stokes flow
equations,
\begin{equation}
\mu\nabla^2\mathbf{u}  =   \boldsymbol{\nabla}p, \quad
\boldsymbol{\nabla}\cdot\mathbf{u} = 0,
\end{equation}
where $\mathbf{u}$ is the fluid velocity field, $p$ the dynamic pressure and
$\mu$ is the dynamic viscosity, hereafter non-dimensionalized to $\mu = 1$. 

\begin{figure}[tbp]	
\begin{center}
\includegraphics[scale=0.8]{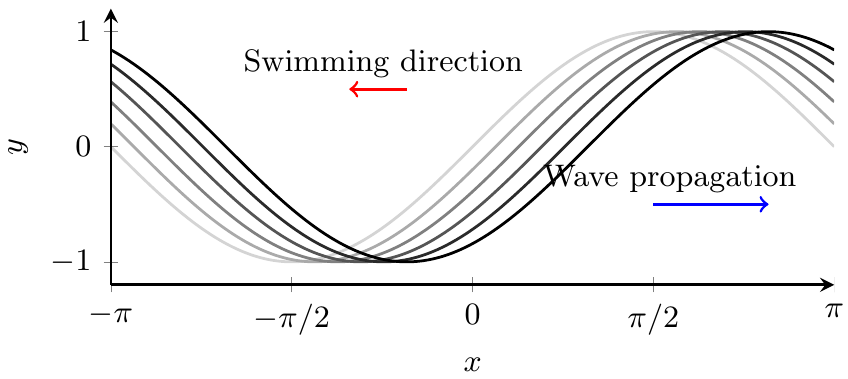}
\end{center}
\caption{A schematic of the wave propagation of a swimming sheet, here
illustrated  by a single sine-wave mode, showing the computational domain ($-\pi$
to $\pi$ along the $x$ axis), direction of wave propagation (in the swimming
frame) and direction of swimming.}	
\label{fig:setup}
\end{figure}

We consider the waving sheet model illustrated in Fig.~\ref{fig:setup}, and
assume the waveform to be fixed and to travel along the positive $x$ direction
at unit speed. The infinite sheet is periodic over the interval $[0,2\pi]$, and
swimming is expected to occur in the $-x$ direction, opposite to the direction
of propagation of the wave \cite{taylor1951analysis}. { Since the
sheet is infinite, there is no extrinsic length-scale to the problem, and as
such we hereafter non-dimensionalise lengths using  the reciprocal of the
wavenumber, $k^{-1}$.} In order for net swimming to take place with no rotation, we
require the wave to be odd about the axis $x = 0$, i.e.~ask that
$y([0,-\pi]{,t}) = -y([0,\pi]{,t})$. Without loss of
generality, the waveform may therefore be described as a Fourier-sine series
where the shape in the swimming frame is described by $y(x,t)$ with 

\begin{equation}
y(x,t) = \sum_{p=1}^P B_p \sin[p(x-t)].
\end{equation}
Even modes for the shape, $B_{2q}$ with $q$ any integer, are always obtained by
our optimization algorithm to be zero, indicating that optimal waveforms are
front-back symmetric waves. { The physical reason underlying this
front-back symmetry is unclear. Due to kinematic reversibility, if the shape was
asymmetric then an equally optimal waveform would be its front-back mirror
image, and thus the optimization procedure would always lead to two symmetric
solutions. This is not the case and a unique, front-back symmetric shape is
always obtained.} We thus consider a general waveform represented by 
\begin{equation}
y(x,t) = \sum_{n=1}^N B_n \sin[(2n - 1)(x-t)],
\end{equation}
and use our computational approach to derive the optimal series of coefficients
{$\{ B_n\}_N, n\leq N$ for increasing N}. { The lack of
an extrinsic length-scale to the infinite sheet means that our choice of first
mode is in some sense arbitrary, and thus we will not consider solutions for
which $|B_1| > 0$, in order to define a fundamental period to the wave.}

To derive the hydrodynamically-optimal waveform we use the standard definition
of swimming efficiency introduced by \citet{lighthill1975mathematical}. We
therefore compare a useful rate of swimming, $\sim U^2$, to the rate of working
of the sheet against the fluid, $W =  -\int_S \mathbf{u} \cdot
\boldsymbol{\sigma}\cdot\mathbf{n} {\,\mathrm{d}s}$, where the fluid stress is
$\boldsymbol{\sigma} = -p\mathbf{I} + (\boldsymbol{\nabla}\mathbf{u} +
(\boldsymbol{\nabla}\mathbf{u})^{T})/2$, {$S$ is the surface of the swimmer over one wavelength},  and  $\mathbf{n}$ is  the unit normal
to the sheet into the fluid. We thus  seek the  set of coefficients $\{B_n\}_N$
that maximize the hydrodynamic efficiency, $\mathcal{E}$, defined as
\begin{equation}
\mathcal{E} = \frac{U^2}{-\int_S \mathbf{u} \cdot \boldsymbol{\sigma} \cdot
\mathbf{n}{\,\mathrm{d}s}},
\label{eq:efficiency}
\end{equation}
and numerically compute the value of the swimming speed, $U$, and surface
stress, $\boldsymbol{\sigma} \cdot \mathbf{n}$. 

In order to impose velocity conditions on the surface of the sheet, we solve the
problem in a frame of reference that moves with the propagating wave. Since the
sheet is stationary in this frame, the velocity of material elements is purely
tangential \cite{lighthill1975mathematical}. By subtracting the normalized wave
speed, this allows us to  retrieve the boundary conditions everywhere along the
sheet as 
\begin{equation}
u(\mathbf{x}) = -Q\cos\theta(\mathbf{x}) + 1, \quad v = -Q\sin\theta(\mathbf{x}),
\end{equation} 
where $Q$ denotes the ratio between the arclength of the waveform in one
wavelength to the wavelength measured along the $x$ direction  and
$\theta(\mathbf{x})$ is the tangent angle of the sheet measured about the $x$
axis. Both the value of $Q$ and the distribution of $\theta$ are functions of
the wave geometry only, and thus of the coefficients $\{B_n\}_N$.


\section{Computational approach} 
\label{sec:numerics}

In order to compute the flow field generated by the sheet, and the resultant
surface stresses, we employ the boundary element method
\cite{youngren1975stokes} with two-dimensional, periodic Green's functions as in
\citet{pozrikidis1987creeping} and \citet{sauzade2011taylor}. At any point
$\mathbf{x}$ along the sheet, the velocity at that point is given by the surface
integral 
\begin{align}
u_j(\mathbf{x}) = \frac{1}{2\pi}\int_S [
&S_{ij}(\mathbf{x}-\mathbf{x}^{\prime})f_i(\mathbf{x}^{\prime}) \nonumber \\
- &T_{ijk}(\mathbf{x}-\mathbf{x}^{\prime})u_i(\mathbf{x}^{\prime})n_k(\mathbf{x}^{\prime})]\,
\mathrm{d}s(\mathbf{x}^{\prime}),
\label{eq:2d_nd_int}
\end{align}
where $\mathbf{n}(\mathbf{x}^{\prime})$ is the unit normal pointing into the fluid at
$\mathbf{x}^{\prime}$ and $f_i = \sigma_{ij}n_j$.  Using the notation $r =
|\hat{\mathbf{x}}|$, the stokeslet tensor,  $S_{ij}$, and stresslet, $T_{ijk}$,
are given by 
\begin{equation} 
S_{ij}(\hat{\mathbf{x}}) = \delta_{ij}\ln r - \frac{\hat{x}_i\hat{x}_j}{r^2},
\quad T_{ijk}(\hat{\mathbf{x}}) = 4\frac{\hat{x}_i\hat{x}_j\hat{x}_k}{r^4},
\label{eq:2d_stokeslet}
\end{equation}
and represent the solution to Stokes flow due to a point force in two dimensions
and the corresponding stress respectively. Since we are modeling an infinite,
$2\pi$-periodic sheet, we have velocity contributions at $\mathbf{x}$ from an
infinite sum of stokeslets and stresslets, 
\begin{equation}
\mathbf{S}^p = \sum_{n = -\infty}^{\infty} \mathbf{I}\ln r_n -
\frac{\hat{\mathbf{x}}_n\hat{\mathbf{x}}_n}{r_n^2}, \quad
\mathbf{T}^p = \sum_{n = -\infty}^{\infty}
4\frac{\hat{\mathbf{x}}_n\hat{\mathbf{x}}_n\hat{\mathbf{x}}_n}{r_n^4},
\end{equation}
where $\hat{\mathbf{x}}_n = \left(x - x^{\prime} + 2\pi n,y - y^{\prime}\right)$
for singularities positioned at $\mathbf{x}^{\prime}$.
These infinite sums may be conveniently expressed in a closed form as
\begin{equation}
\begin{array}{rl}
S^p_{xx} &= A + \hat{y}\partial_{\hat{y}}A - 1, \\
S^p_{yy} &= A - \hat{y}\partial_{\hat{y}}A, \\
S^p_{xy} &= -\hat{y}\partial_{\hat{x}}A = S^p_{yx},\\ 
T^p_{xxx} &= 2\partial_{\hat{x}}(2A + \hat{y}\partial_{\hat{y}}A),\\
\end{array} \qquad
\begin{array}{rl}
T^p_{xxy} &= 2\partial_{\hat{y}}(\hat{y}\partial_{\hat{y}}A),\\
T^p_{xyy} &= -2\hat{y}\partial_{\hat{x}\hat{y}}A,\\
T^p_{yyy} &= 2(\partial_{\hat{y}}A - \hat{y}\partial_{\hat{y}\hat{y}}A),\\
T^p_{ijk} &= T^p_{kij} = T^p_{jki},
\end{array}
\end{equation}
where $A = \frac{1}{2}\ln\left[2\cosh(\hat{y}_0) - 2\cos(\hat{x}_0) \right]$.
Note that these are equivalent to, but differ by a minus sign from, those found in
\citet{sauzade2011taylor}, due to our adoption of the sign convention from
\citet{pozrikidis1987creeping}.

For the computational procedure, the sheet is discretized into $500$ straight
line segments of constant force per unit length, i.e.~the components $f_{1,2}$
in equation \eqref{eq:2d_nd_int} are constant over each straight line element.
This discretization breaks equation \eqref{eq:2d_nd_int} into a sum of line
integrals of singularities, multiplied by the unknown force per unit length.
Numerical evaluation of each non-singular line integral is performed with
four-point Gaussian quadrature, whilst singular integrals are treated
analytically. This numerical discretization produced a $< 0.02\%$ relative
difference in the calculated swimming velocity and efficiency for both kinked
and unkinked example sheets when compared to simulations with $600$ and $800$
elements with $8$- and $16$-point Gaussian quadrature, whilst still allowing
calculation in a reasonable time. By decoupling the numerical quadrature from
the force discretization, comparable accuracy is achieved for relatively smaller
linear systems \cite{smith2009boundary}. The computational mesh is refined
locally around regions of high curvature appearing as a result of the
optimization in order to resolve potential kinks and singular shapes. Using
these parameters, we find convergence of the waveform and optimal efficiency for
$N=30$. In order to reach this number of coefficients quickly, it is important
to provide a good starting guess for the coefficients $B_n$ at the beginning of
the optimization procedure.  Since we are interested in the convergence of our
waveform with an increased number of points, we numerically optimize for $n =
1,\dots,30$ sequentially, and use the converged optimal coefficients for the
$n=N-1$ waveform as an initial guess for the $n=N$ waveform, with $B_N$
initially set to zero.  { In order to ensure that our solution
represents the global maximum of the hydrodynamic efficiency
\eqref{eq:efficiency}, multiple initial conditions were tried, all found to be
optimized to the same waveform. Optimization is carried out using the standard
\textsf{fminunc} function in \textsf{Matlab} using the \textsf{quasi-newton}
algorithm.}


\section{The shape of the optimal swimming sheet}
\label{sec:results}

\begin{figure*}[tbp]	
\begin{center}
\includegraphics[scale=0.8]{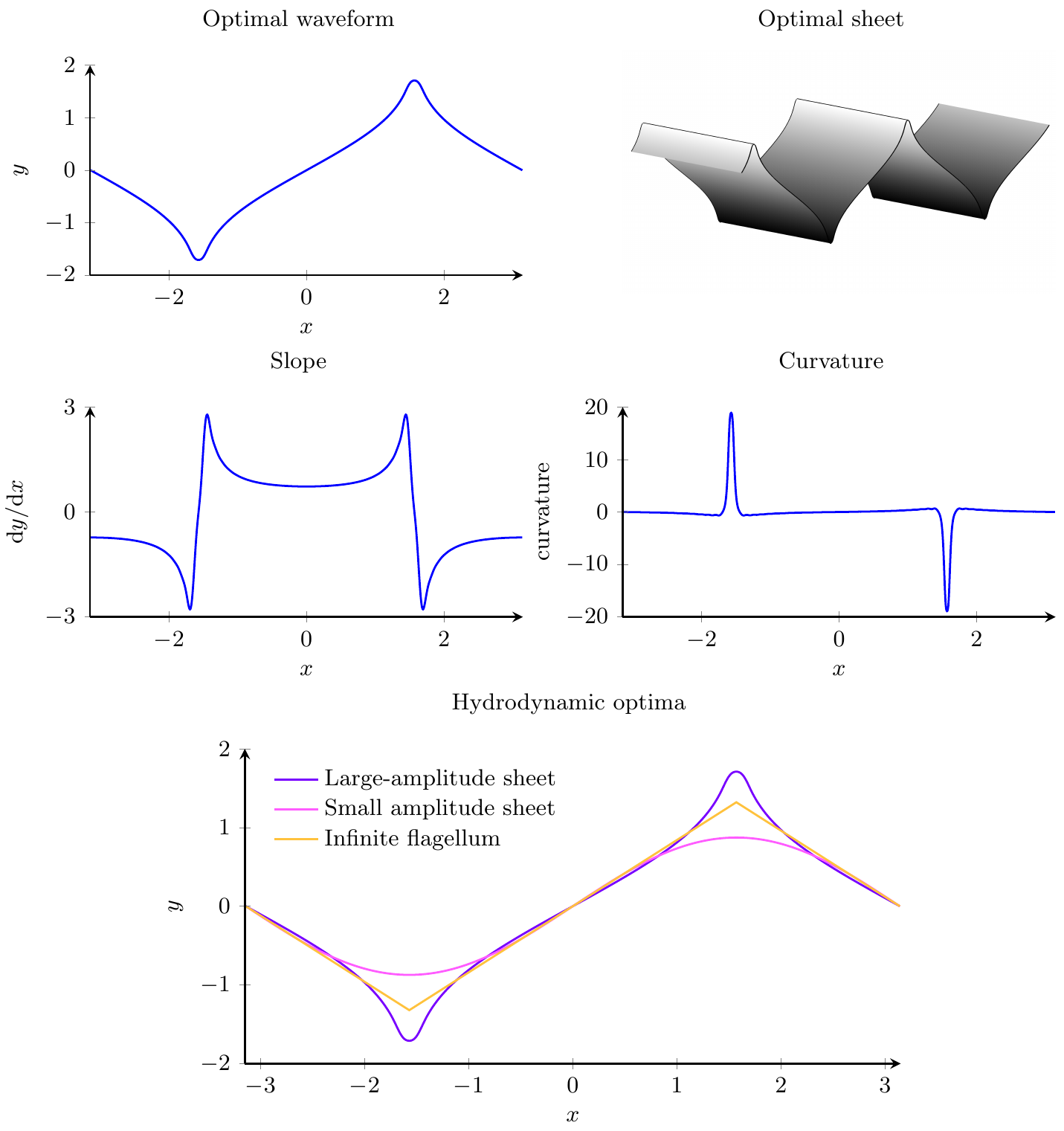}
\end{center}
\caption{The optimal swimming sheet. (a) Optimal waveform obtained
computationally, which exhibits an almost straight region followed by a
steepening of the wave crest into a rounded cap; (b) Three-dimensional
visualization of the optimal swimming sheet;  Distribution of slope (c) and
curvature (d) of the optimal waveform showing largely straight regions; 
{(e) Waveform of the optimal sheet shown for comparison with the pure sine-wave
predicted by small amplitude theory \cite{taylor1951analysis} and Lighthill's optimal triangle wave for a
flagellum with drag anisotropy ratio of $1/2$ \cite{lighthill1975mathematical}.}}	
\label{fig:optim_form}
\end{figure*}

With this framework established, we are able to compute the shape of the optimal
swimming sheet. It is first instructive to ask what would be predicted
from the analytical small-amplitude approach. In that case, the waveform is
written as
\begin{equation}
y(x,t) = \epsilon \sum_{n = 1}^{\infty} B_n e^{in(x - t)},
\end{equation}
where $\epsilon$ is a small dimensionless amplitude. To second order in the amplitude \cite{taylor1951analysis},
the swimming speed, $U$, and work done against the fluid, $W$, are given by
\begin{equation}
U \sim \sum_{n=1}^{\infty} n^2|B_n|^2, \quad W \sim \sum_{n=1}^{\infty} n^3|B_n|^2.
\end{equation}
Because the work done is proportional to $n^3$, whereas the swimming velocity is
proportional to $n^2$, higher-order modes are energetically penalized compared
to lower modes. Therefore for a fixed amount of mechanical power expended by
the swimmer, it is more efficient to distribute all of that power to the first
mode $n=1$; under the small amplitude approximation, the most efficient waveform
is therefore a single sine-wave of period $2\pi$.

Relaxing the small-amplitude constraint, we show in Fig.~\ref{fig:optim_form}a
the optimal waveform obtained numerically for $N = 30$ odd Fourier modes. The
set of coefficients $B_n$ that describe this waveform are given in  Appendix
\ref{sec:appendix} and a three-dimensional sheet propagating this wave is
further shown in Fig.~\ref{fig:optim_form}b. The optimal swimming sheet appears
to take the shape of a regularized cusp wave, qualitatively different from the
single-mode sine-wave predicted by the asymptotic analysis. We further plot in
Fig.~\ref{fig:optim_form}c the distribution of slopes along the sheet, showing
that the  wave has an almost straight section (constant slope), steepening
towards smooth wave crest. The angle of the slope at the point of symmetry $x =
0$ is approximately $ 36.1^{\circ}$, close to the optimum value of
$40.06^{\circ}$  obtained for Lighthill's three-dimensional flagellum via
resistive-force theory with a drag anisotropy ratio of $1/2$
\cite{lighthill1975mathematical}. We display in Fig.~\ref{fig:optim_form}d the
distribution of curvature along the sheet; while the curvature at the wave crest
increases, it remains finite. { For comparison, our predicted optimal waveform is plotted against the pure sine-wave of small amplitude theory \cite{taylor1951analysis} and the triangle wave predicted by Lighthill \cite{lighthill1975mathematical} in Fig.~\ref{fig:optim_form}e.} 

\begin{figure}[tbp]
\begin{center}
\includegraphics[scale=0.8]{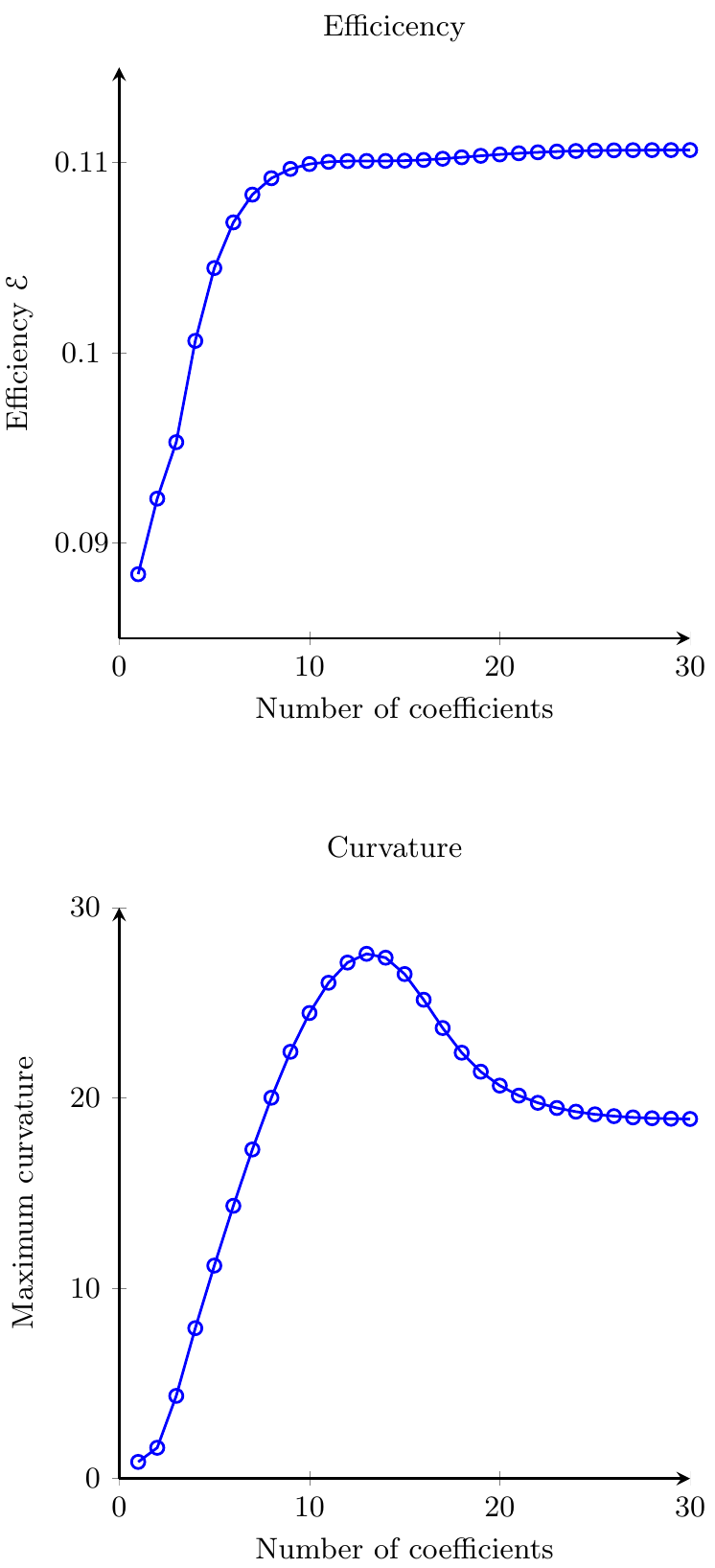}
\end{center}
\caption{Convergence of the swimming efficiency (a) and maximum wave curvature
(b) as a function of the number of odd Fourier modes in the optimisation, $N$.}	
\label{fig:eff_curv}
\end{figure}

Since our optimization procedure finds the optimal solution for incremental
values of the number of coefficients, $N$, used to describe the wave, we can
investigate convergence of all optimal waveforms described by $N $ ranging from
1 to 30. The convergence for the swimming efficiency  is shown in
Fig.~\ref{fig:eff_curv}a while the dependence of the maximum curvature on $N$
is plotted in Fig.~\ref{fig:eff_curv}b.
\begin{figure}[t]
\begin{center}
\includegraphics[scale=0.8]{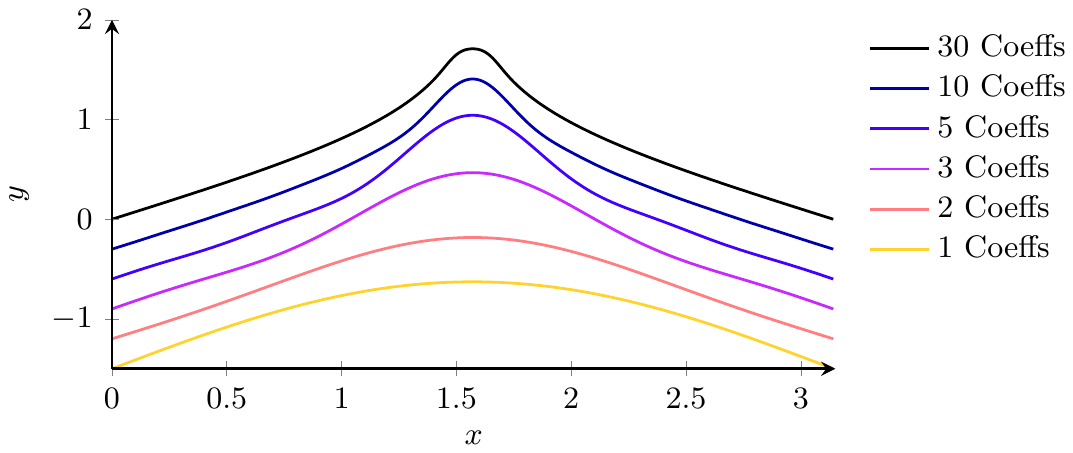}
\end{center}
\caption{Convergence of the shape of the optimal waveform for increasing numbers
of odd coefficients. Shapes are shifted by $0.1\pi$ along the $y$ direction for
visualization.}	
\label{fig:slope_conv}
\end{figure}
For $N=30$, the optimal waveform is over $25\%$ more efficient than the optimal
one-mode sine-wave ($N=1$). The swimming efficiency appears to reach its
asymptote near $N=13$, which corresponds to the peak in the maximum curvature,
but thereafter continues to increase slightly before reaching its converged
value of $\mathcal{E} \approx 0.11065$. This slight increase is accompanied by a
decrease in the  maximum curvature of the optimal waveforms for $N \geq 14$.
Up to $N = 13$, it appears that subsequent modes serve to steepen the waveform
as it approaches around the crest. Such steepening is likely hydrodynamically favorable
in two dimensions since fluid cannot pass around the sheet as it would around
three-dimensional flagellum. However, steepening results in a region of high
curvature at the wave crest, which induces locally high viscous dissipation in
the fluid, and so there appears to be an efficiency trade-off between wave steepening and
minimizing curvature. For $N \geq 14$, the wavelength of the Fourier modes is on
the order of the length of the cap on the wave crest. These modes are then able
to decrease the maximum curvature without decreasing the slope of the wave,
yielding small increases in efficiency until the curvature converges for $N \geq
30$. We further display the  convergence of the optimal waveform as a function
of the  number of coefficients, $N$,  in Fig.~\ref{fig:slope_conv}. Despite the
decrease in maximum curvature seen in Fig.~\ref{fig:eff_curv}b for $N \geq 14$,
all waveforms between $N = 10$ and $30$ are virtually indistinguishable by eye.

\begin{figure}[t]
\begin{center}
\includegraphics[width=.5\textwidth]{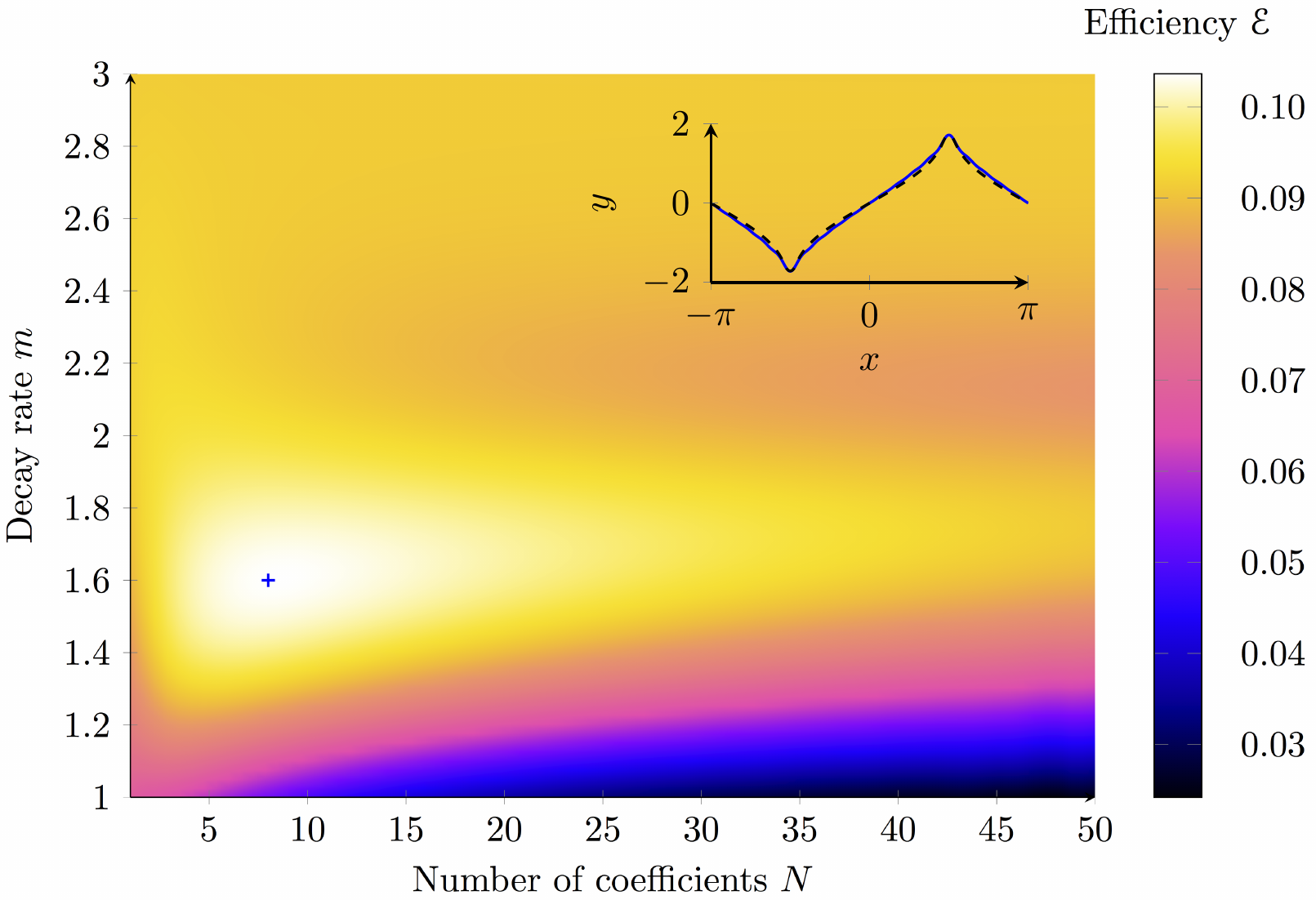}
\end{center}
\caption{Efficiences of optimal truncated `cusp-like' waveforms
\eqref{eq:cusp_wave} as a function of number of odd modes in the description $N$
and decay rate $m$, showing a maximum at $N = 9, m \approx 1.6$ for regularized
cusps. The waveform corresponding to this maximum is shown inset (blue, solid)
in comparison to the full optimal of fig.~\ref{fig:optim_form}a (black, dashed),
showing good qualitative agreement.}	
\label{fig:cusp_eff}
\end{figure}

The trade-off between wave steepening and reducing curvature can be further investigated by
examining a family of waves of the form
\begin{equation}
B_n =C \frac{(-1)^{n-1}}{(2n - 1)^m}, \,\,n\geq 1,
\label{eq:cusp_wave}
\end{equation}
where $C$ is a constant. The value of $m$ dictates the decay of the Fourier
coefficients with the Fourier mode, and with the choice $m = 2,$
Eq.~\eqref{eq:cusp_wave} leads to the triangular waveform of Lighthill's optimal
flagellum. The choice of alternating sign is informed both by the series for the
triangle wave, and by the coefficients of our optimal solution (up to $N = 14$).
Cusps are obtained for $m < 2$ and rounded-off waves for $m > 2$ and, by
truncating the series at small values of $N$, we retrieve an approximate
regularized cusp wave. Fig.~\ref{fig:cusp_eff} shows iso-contours of the efficiency of waves
described by Eq.~\eqref{eq:cusp_wave} for the optimal value of the amplitude,
$C$, as a function of the number of coefficients used to describe the wave, $N$,
and the decay rate, $m$. The optimal efficiency $\mathcal{E} = 0.1037$ of such
waves occurs when $N = 9$, for $C = 1.237$ and $m = 1.609$, which corresponds to
a slower decay of the fourier modes than Lighthill's wave. The waveform
associated with this optimal is plotted inset in Fig.~\ref{fig:cusp_eff} (blue,
solid), showing a strong similarity to our fully converged optimal computed for
$30$ coefficients (black, dashed).  

If a large enough number of odd modes ($N \geq 50$) is used to describe the
curve, the kink at the wave crest is sufficiently resolved as to no longer be
regularized. In this case, the optimal jumps to an unkinked profile with which
$C = 0.9173$ and $m = 2.923$, yielding an efficiency of just $0.0912$ and
demonstrating the detrimental effect of kinked waveforms on hydrodynamic
efficiency when non-local effects are taken into account. The optimal within
this family is thus more efficient than any kinked wave. Furthermore, this
result suggests that by fully resolving the hydrodynamics around Lighthill's
optimal flagellum, viscous dissipation associated with the kink might also
regularize this waveform.


\section{Discussion}
\label{sec:discussion}

Taylor's swimming sheet model is commonly used to address a range of phenomena
in the biological physics of small-scale locomotion. A natural question to raise
is the relevance of a two-dimensional geometry to the three-dimensional
locomotion of flagellated cells. In this paper, we used the boundary element
method to compute the swimming efficiency of arbitrary waveforms in two
dimensions. By focusing on the question of optimal waveform for locomotion,
{{we show that the optimal two-dimensional waveform is a regularized
cusp, which is about 25\% more efficient than a simple sine-wave.}}
{{This result is different from the three-dimensional
hydrodynamically-optimal triangle wave derived by Lighthill
\cite{lighthill1975mathematical}; the slope of the straight section is
shallower, the waveform steepens towards the wave crest and there is no
discontinuity in the slope but rather a regularized cusp. The result is also
different to the three-dimensional internally-optimal wave, which is composed of
circular arcs joined by straight lines \cite{lauga2013shape}}}. Although it is
know that the dynamics of a swimming sheet can provide qualitative insight into
the hydrodynamics of small-scale locomotion, {differences} with
three-dimensional results exist therefore at large amplitude. 
 

\subsection*{Acknowledgements}
\label{subsec:acknowledgements}
The authors would like to thank Gwynn Elfring for useful discussions.  
This work was funded in part by the European Union through a Marie Curie CIG to EL.

\begin{appendix}
\section{Coefficients of the optimal waveform}
\label{sec:appendix}
The Fourier coefficients for the optimal waveform for $N = 30$ are given by
$\times 10^{-2}:$
\begin{align}
B_1    &= 114.6    &B_2    &= -24.94   &B_3    &= 12.05    \nonumber \\   
B_4    &= -7.090   &B_5    &= 4.534    &B_6    &= -3.017   \nonumber \\
B_7    &= 2.039    &B_8    &= -1.378   &B_9    &= 0.9180   \nonumber \\  
B_{10} &= -0.5925  &B_{11} &= 0.3611   &B_{12} &= -0.1973  \nonumber \\ 
B_{13} &= 0.08291  &B_{14} &= -0.00511 &B_{15} &= -0.04558 \nonumber \\
B_{16} &= 0.07617  &B_{17} &= -0.09207 &B_{18} &= 0.09738  \nonumber \\
B_{19} &= -0.09531 &B_{20} &=  0.08831 &B_{21} &= -0.07831 \nonumber \\
B_{22} &=  0.06675 &B_{23} &= -0.05475 &B_{24} &= 0.04310  \nonumber \\
B_{25} &= -0.03239 &B_{26} &= 0.02299  &B_{27} &= -0.01515 \nonumber \\
B_{28} &= 0.008958 &B_{29} &= -0.004427&B_{30} &= 0.001472 \nonumber
\end{align}
\end{appendix}

\bibliography{my_refs}

\end{document}